\documentclass[lettersize,magazine]{IEEEtran}
\usepackage{amsmath,amsfonts}
\usepackage{algorithmic}
\usepackage{algorithm}
\usepackage{array}
\usepackage[caption=false,font=footnotesize,labelfont=rm,textfont=rm]{subfig}
\usepackage{multirow}
\usepackage{textcomp}
\usepackage{stfloats}
\usepackage{xcolor}
\usepackage{url}
\usepackage{verbatim}
\usepackage{graphicx}
\usepackage{cite}
\usepackage{ragged2e}
\begin{document}
\bstctlcite{setting}
\title{Toward Trustworthy Digital Twins in \textcolor{black}{AI Agent}-based Wireless Network Optimization: Challenges, Solutions, and Opportunities}
\author{Zhenyu~Tao,~\IEEEmembership{Graduate Student Member,~IEEE},
        Wei~Xu,~\IEEEmembership{Fellow,~IEEE},
        and~Xiaohu~You,~\IEEEmembership{Fellow,~IEEE}

\thanks{Zhenyu Tao, Wei Xu, and Xiaohu You are with the National Mobile Communications Research Lab, Southeast University, Nanjing 210096, China, and also with the Pervasive Communication Research Center, Purple Mountain Laboratories, Nanjing 211111, China (email: \{zhenyu\_tao, wxu, xhyu\}@seu.edu.cn). \textit{Corresponding author: Xiaohu You.}}
}

\markboth{IEEE Wireless Communications}
{Shell \MakeLowercase{\textit{et al.}}: A Sample Article Using IEEEtran.cls for IEEE Journals}


\maketitle

\begin{abstract}
Optimizing modern wireless networks is exceptionally challenging due to their high dynamism and complexity. While the \textcolor{black}{AI agent} powered by reinforcement learning (RL) offers a promising solution, its practical application is limited by prohibitive exploration costs and potential risks in the real world. The emerging digital twin (DT) technology provides a safe and controlled virtual environment for \textcolor{black}{agent} training, but its effectiveness critically depends on the DT's \textcolor{black}{reliability}. Policies trained in \textcolor{black}{an unreliable} DT that does not accurately represent the physical network may experience severe performance degradation upon real-world deployment. In this article, we introduce a new DT evaluation framework to ensure trustworthy DTs in \textcolor{black}{AI agent}-based network optimization. This framework shifts from \textcolor{black}{model-level accuracy}, such as wireless channel and user trajectory similarities, to a more holistic, task-centric DT assessment\textcolor{black}{, which relies on the Markov decision process that the agent actually perceives.} We demonstrate it as an effective guideline for design, selection, and lifecycle management of wireless network DTs. A comprehensive case study on a real-world wireless network testbed shows how this evaluation framework is used to pre-filter candidate DTs, leading to a significant reduction in training and testing costs without sacrificing deployment performance. Finally, potential research opportunities are discussed.
\end{abstract}

\section{Introduction}
The vision for the sixth-generation (6G) wireless networks promises to support a new wave of transformative applications, from immersive extended reality (XR) to the massive-scale Internet of things (IoT). However, realizing this vision presents unprecedented challenges for network optimization~\cite{you2023toward}. Firstly, 6G networks are expected to accommodate various services, each with stringent and diversified performance requirements on latency, throughput, and reliability. This creates an environment of significant dynamism, driven by varying traffic, diversified user mobility, and rapidly changing channel conditions. Meanwhile, the intricate and heterogeneous 6G network architecture leads to optimization tasks with incredibly high dimensionality. For a typical network optimization task like resource allocation, a network controller should adjust hundreds of parameters simultaneously, including resource block assignment, beamforming, and power management, to seek optimal overall performance. 

The combined challenges of dynamic and high-dimensional network environments surpass the capabilities of classical network optimization techniques, such as linear programming-based resource allocation~\cite{Elsherif2015}. These methods often rely on precise mathematical models that are difficult to formulate in real-world 6G networks and struggle to find real-time solutions due to prohibitive computational complexity. In response, research has shifted towards \textcolor{black}{AI agent}-empowered solutions, with reinforcement learning (RL) emerging as a powerful, data-driven framework~\cite{DRLsurvey}. By directly interacting with network environments to refine control policies parameterized by neural networks (NNs), ranging from simple multi-layer perceptrons (MLPs) to complex large language models, RL is uniquely suited to handle complex and ever-changing 6G network optimization tasks~\cite{DRLsurvey}.

However, the foundational trial-and-error learning process of RL is fraught with peril if conducted on a live operational network. Unrestricted exploration in the real world could lead to prohibitive costs, significant degradation in quality of service (QoS), potential service level agreement (SLA) violations, and even risks to network stability~\cite{DRLsurvey}. This practical barrier has facilitated the digital twin (DT)-driven RL paradigm~\cite{DTRL}. By creating a virtual replica of the physical network, a DT provides a safe and efficient environment for \textcolor{black}{agent} training~\cite{Nvidia}. For instance, in resource allocation tasks, an RL agent can explore countless strategies rapidly within such a DT, assigning different resource blocks and power levels to simulated users under various traffic and channel conditions, all without affecting the operation of physical networks. After being exhaustively trained and validated, the resulting agent policy can then be deployed to real-world network controllers, thereby avoiding dangerous exploration~\cite{taojsac}, reducing trial-and-error costs~\cite{ruah2023}, and accelerating the entire training process~\cite{zhang2023digital}.

Despite these compelling advantages, the effectiveness of the DT-driven RL paradigm closely depends on the DT's fidelity to the real world. While a precise DT can yield highly effective control policies, an inaccurate one would result in a risky simulation-to-reality gap~\cite{ruah2025}, leading to significant performance degradation upon policy deployment. An RL agent trained in a DT with an oversimplified, stable traffic model learns a rigid resource allocation policy that may fail upon real-world deployment due to bursty and dynamic traffic in physical networks. This raises the pivotal question of this article: \textit{How can we trust an AI agent trained in a DT to perform effectively and safely in real wireless networks?}

\begin{figure*}[!t]
\centering
\includegraphics[width=1\textwidth]{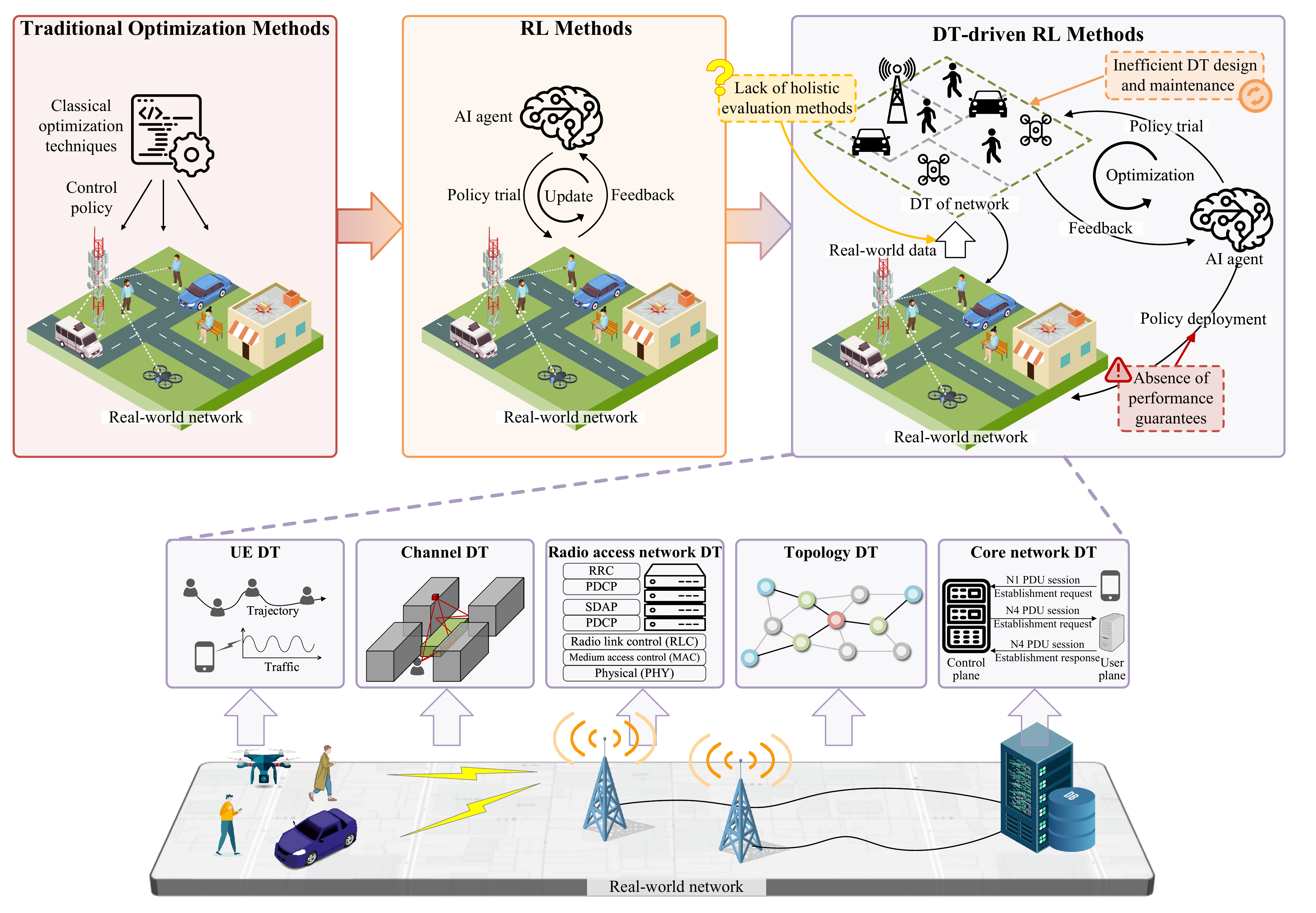}
\caption{\textcolor{black}{Evolution of wireless network optimization techniques.}}
\label{fig1}
\end{figure*}

Addressing this question requires a solid bridge between DT and real physical world, the establishment of which poses three critical challenges:
\begin{itemize}
    \item \textbf{Lack of holistic evaluation methods:} A DT for network optimization is typically a composite system, integrating multiple models of physical entities, such as user equipment~\cite{taojsac}, wireless channel~\cite{Nvidia}, and core network, as schematized in Fig.~\ref{fig1}. Current assessment methods are designed to evaluate the physical accuracy of each model~\cite{InternetDT}, e.g., channel model accuracy. However, the effectiveness of a DT for \textcolor{black}{agent} training is a holistic property arising from interactions among all its constituent components, which is hardly captured by straightforward aggregations of individual \textcolor{black}{model accuracies}. 
    \item \textbf{Inefficient DT design and maintenance:} The absence of an effective evaluation metric forces DT development to become a process driven by empiricism and expert intuition rather than principled refinement. How do we choose the most suitable models for DT construction? How much data is sufficient to ensure its \textcolor{black}{reliability}? When should the DT be updated to reflect changes in the real world? These unresolved questions lead to either wasted resources in building an overly complex DT or poor performance from an inadequate one.
    \item \textbf{Absence of performance guarantees:} Critically, there is no reliable way to predict how a policy trained in the DT will perform upon deployment into the physical wireless network~\cite{DTsurvey}. If we cannot theoretically analyze and guarantee its real-world performance before actual deployment, the DT's core advantage of avoiding risky real-world exploration is fundamentally undermined.
 \end{itemize}

In this article, we provide a holistic DT evaluation framework as a theoretically guaranteed solution to ensure trustworthy DTs in \textcolor{black}{AI agent}-based network optimization, detailed in Section~II. Section~III presents potential applications by integrating such evaluation into DT-driven network optimization. Following this, Section~IV provides a real-world case study validating the framework's practical value. Finally, we conclude with promising future directions.

\section{\textcolor{black}{Holistic DT Evaluation as a Solution}}
To establish a bridge between DT and the real world for wireless network optimization, a new evaluation paradigm is required that \textcolor{black}{advances from model-level accuracy to holistic, task-centric fidelity from the agent's perspective}.

\subsection{\textcolor{black}{Moving Beyond Model-Level Accuracy}}
\begin{figure*}[!t]
\centering
\includegraphics[width=1\textwidth]{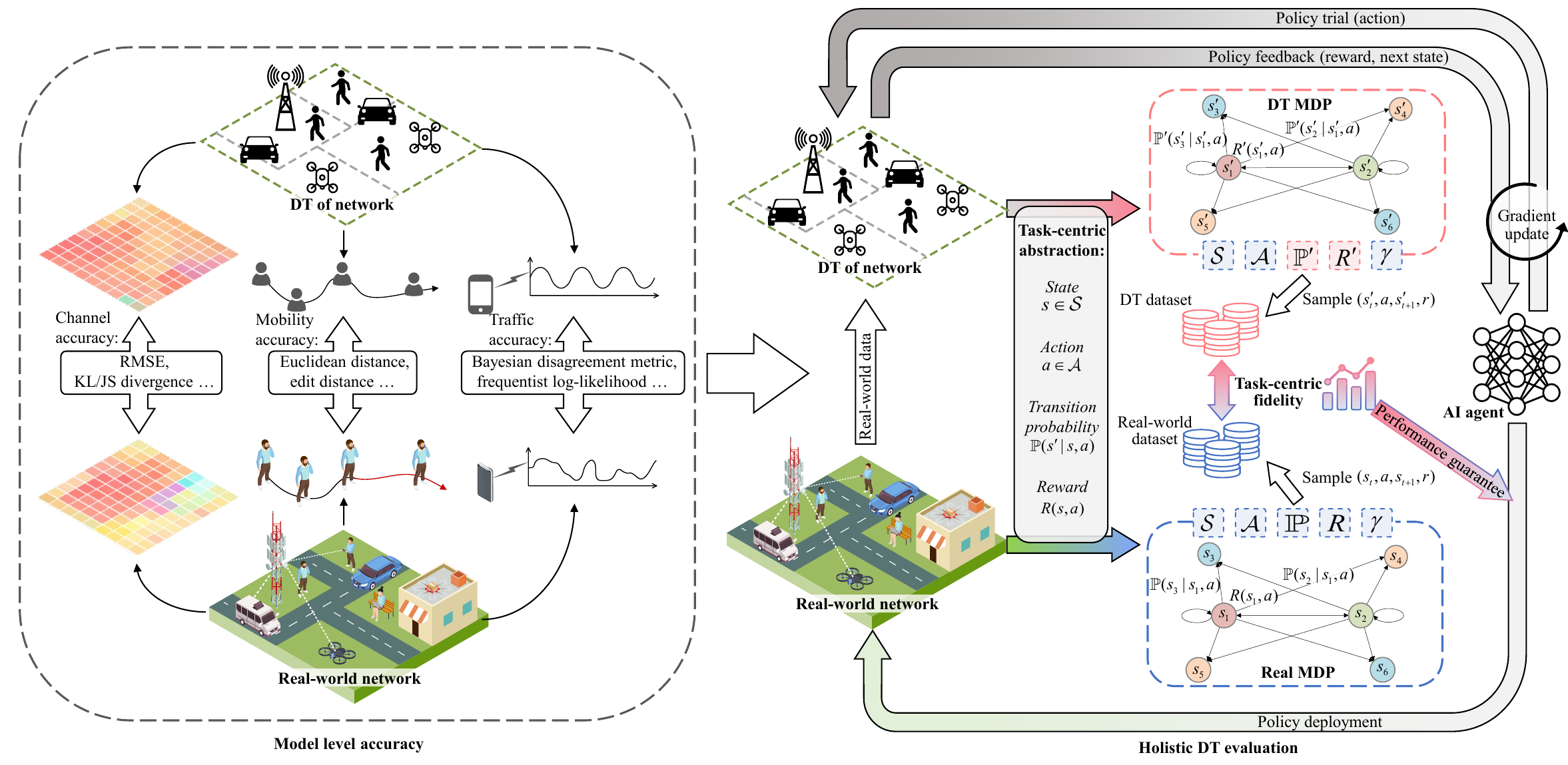}
\caption{\textcolor{black}{Holistic DT evaluation framework: \textbf{(1)} define the agent-perspective MDP for both the real network and the DT according to a specific optimization task; \textbf{(2)} sample (state, action, reward, next-state) tuples from each environment under its existing operational policy; \textbf{(3)} estimate the two MDPs from data samples; \textbf{(4)} compute the task-centric fidelity and derive the performance guarantee on policy deployment.}}
\label{fig2}
\end{figure*}

\textcolor{black}{Conventional DT evaluation approaches are model-level}, focusing on the physical accuracy of \textcolor{black}{each constituent model} in DT. As shown in Fig.~\ref{fig2}, a channel DT's \textcolor{black}{accuracy} is often measured using structural similarity or root mean squared error (RMSE) against real-world data~\cite{Nvidia}. Likewise, user mobility models can be evaluated by the Euclidean distance or edit distance between their simulated trajectories and real ones~\cite{taojsac}, while traffic models are often assessed using Bayesian disagreement or frequentist log-likelihood prediction metrics~\cite{ruah2023}. \textcolor{black}{We refer to such per-component, physically absolute, and agent-agnostic metrics collectively as \emph{model-level accuracy}.}

While useful for evaluating individual predictive models, these \textcolor{black}{model-level} approaches are fundamentally insufficient for validating a DT's effectiveness for \textcolor{black}{AI agent} training. They fail to capture the \textcolor{black}{behavioral mismatch} between the two environments, as the RL training is associated with complex interactions between DT components, not just isolated model accuracies. \textcolor{black}{For instance, consider two DTs with the same average per-model accuracy: one combines a high-accuracy channel model with a low-accuracy mobility model, while the other uses two moderately accurate models. Conventional metrics that aggregate isolated model accuracies can neither distinguish such DTs nor predict their agent training efficacies. This motivates evaluation methods from the agent's perspective, which is termed as \emph{task-centric fidelity}.}

\subsection{\textcolor{black}{Holistic DT Evaluation via Task-Centric Fidelity}} 
\textcolor{black}{For a wireless network optimization task, the agent perceives the dynamic real network as a Markov decision process (MDP). Regardless of the specific task, the agent's view of the environment is captured by the same four elements:}
\begin{itemize}
    \item \textbf{State $s\!\in\!\mathcal{S}$:} The network information used by the RL agent to make a control decision, e.g., current base station loads, user traffic demands.
    \item \textbf{Action $a\!\in\!\mathcal{A}$:} The operation performed by the agent, e.g., a specific resource block assignment scheme.
    \item \textbf{Transition probability $\mathbb{P}(\tilde{s}|s,a)$:} The underlying dynamics of the environment that govern how a state and an action result in a new state $\tilde{s}$.
    \item \textbf{Reward function $R(s,a)$:} The immediate outcome of an action, associated with the optimization goal, e.g., the resulting overall network throughput.
\end{itemize}
\textcolor{black}{This MDP view applies equally to all AI agents. A traditional RL agent models its state and action as pre-defined structured vectors, while LLM-based agentic AI maps these directly to its textual interface. Its state is the input context, such as the prompt in AgentRAN~\cite{AgentRAN} containing the agent's role, recent KPIs, and decision history, and its action is the generated control decision.} \textcolor{black}{The DT used for agent training is regarded as a similar MDP that aims to mimic the real network.}

\textcolor{black}{We define \emph{task-centric fidelity} as how closely these two MDPs behave from the agent's perspective. Concretely, given a specific network optimization task, we abstract the real network and the DT into their respective MDPs according to the task definition, and evaluate the inter-MDP similarity. A higher fidelity means the agent perceives the DT and the real-world network as more equivalent, and therefore translates into a smaller performance loss when a policy trained in the DT is deployed on the real network.}

\textcolor{black}{To access these MDPs in practice, since their analytical forms are typically unavailable in the real world,} we collect data samples from both environments by monitoring the running real network and the DT. \textcolor{black}{Rather than recording every precise physical value such as signal strength or user location, we log only the information the agent itself sees during learning: the states it receives, the actions it takes, and the resulting rewards.} For instance, in a resource allocation task, we record the network status used to make each allocation decision together with the resulting overall throughput as the reward. \textcolor{black}{Because transitions and rewards are inherent properties of the environment and are independent of the control policy, this sampling can be carried out under the network's existing operational policy, for example, FIFO or priority-based admission control and round-robin or proportional-fair scheduling, without injecting additional exploration actions or degrading QoS. The collected (state, action, reward, next-state) tuples then yield data-driven approximations of the two MDPs for task-centric fidelity calculation. Finally, the resulting fidelity score can be directly linked to a performance guarantee on the real-world deployment of any policy trained in the DT, which is crucial for ensuring the reliability of the DT-driven RL paradigm. The entire process is illustrated in Fig.~\ref{fig2}.}

\subsection{\textcolor{black}{DT-BSM: A Tractable Instantiation}}
\textcolor{black}{The DT-bisimulation metric (DT-BSM)~\cite{taotsp} is a scalar score that quantifies agent-perceived MDP mismatch between the DT and the real network, and serves as a tractable instantiation of the task-centric fidelity introduced in Sec.~II.B. It answers a question: \emph{if the same actions are taken in the two environments, how large can the long-term reward gap become?} A lower DT-BSM value therefore certifies a higher task-centric fidelity, instead of physical accuracy of individual DT components.}

\textcolor{black}{Specifically, DT-BSM captures long-term discrepancies in both $P$ and $R$ through the Wasserstein distance~\cite{transport}, which is well suited to sparse and possibly disjoint state transitions typical in DT settings. It is computed on the two estimated MDPs, without interfering with the operating network, and the minimum complexity~\cite{taotsp} scales as $\mathcal{O}(|\mathcal{A}||\mathcal{S}|^2)$. For LLM-based agentic AI, a naive token-level partitioning of the prompt is impractical, since semantically equivalent prompts may differ in wording; instead, the LLM itself can be used to summarize each prompt into a compact latent embedding, and the calculation is run in this much smaller latent state space. For tasks with an excessively large state space, a state abstraction technique can be applied to merge similar states before calculation, further reducing the cost. }

\textcolor{black}{As a tractable path to task-centric fidelity, DT-BSM provides a direct, provable connection to deployment performance. It establishes an upper bound on the deployment performance loss by a linear summation of agent training loss and the DT-BSM mismatch value, finally providing a quantitative and theoretically grounded guarantee of the real-world performance of the DT-trained agent policy.}

\section{Enhancing Network DTs Through Evaluation}
The holistic DT evaluation framework unlocks significant opportunities to enhance the entire lifecycle of wireless network DTs, from their initial construction to their long-term maintenance. In this section, we explore how evaluation can be applied to two primary categories of DTs in wireless network optimization: simpler, MDP-level DTs that directly model system dynamics, and more complex, system-level DTs that consist of multiple wireless network components.

\subsection{Guiding the Construction of MDP-level DTs}

\begin{figure*}[!t]
\centering
\includegraphics[width=0.9\textwidth]{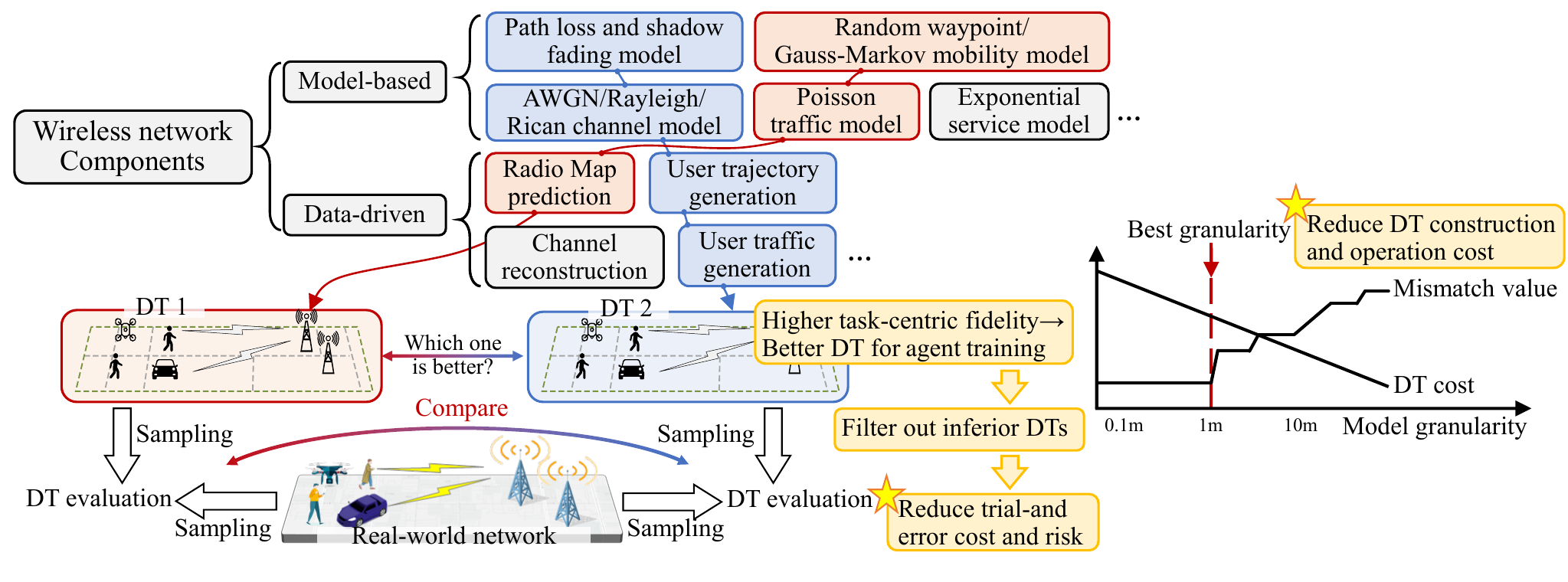}
\caption{\textcolor{black}{Enhancing system-level DT through evaluation.}}
\label{fig3}
\end{figure*}

MDP-level DTs are those that directly seek to model the transition probabilities $\mathbb{P}$ and reward functions $R$ in real-world wireless networks~\cite{zhang2023digital}. DT evaluation provides crucial guidance for creating such models via two main approaches: statistical estimation from collected samples and approximation using NNs.

\textbf{Sampling-based DT:} A straightforward method to build a DT is to collect a massive dataset of state-transition tuples, i.e., state, action, reward, and next state, from the real-world network and use it to statistically estimate the underlying MDP. This approach, however, inevitably leads to a critical question: How much data is enough to ensure the DT is reliable? Collecting too little data results in an inaccurate DT, while collecting too much incurs unnecessary costs. Here, the DT evaluation framework provides a quantitative solution. For instance, DT-BSM offers explicit sample complexity to calculate the minimum data required to achieve a target accuracy level~\cite{taotsp}. This transforms data collection from an open-ended endeavor into a targeted process with a clear stopping criterion, ensuring that resources are used efficiently to build a DT that is provably good enough for subsequent RL applications.

\textbf{NN-based DTs:} For systems with vast state spaces, exhaustive sampling is infeasible. In such cases, a common approach is to use NNs to learn a parameterized model of environment dynamics from sparse data~\cite{zhang2023digital}. However, conventional NN training only captures one-step dynamics, as it uses cross-entropy loss to predict transition probabilities and mean squared error to predict rewards. This approach could be effective on the training set but often leads to overfitting, producing a model that is predictively accurate but behaviorally flawed for RL agents. In contrast, the holistic evaluation of DT provides a multi-step assessment of the model's \textcolor{black}{task-centric fidelity} from an RL perspective. By integrating such evaluation into the DT training loop, either as a metric to guide hyperparameter tuning or directly as a regularization term in the loss function, we can ensure that the learned DT model captures the long-term dynamics essential for decision-making, thereby avoiding overfitting and underfitting, and finally yielding a reliable DT.

\subsection{Optimizing Complex System-level DTs}

In most applications in practice, a wireless network DT is not a singular model but a complex, system-level simulator composed of multiple wireless network components, for instance, an integration of a channel model~\cite{Nvidia}, a mobility model~\cite{taojsac}, and a traffic model~\cite{ruah2023}. The selection and configuration of these components, which we refer to as DT orchestration, play a crucial role in determining the \textcolor{black}{accuracy} and effectiveness of DT for specific RL tasks. In such cases, task-centric DT evaluation would effectively guide its orchestration.

\textbf{Evaluation-guided DT orchestration:} Typically, DT orchestration faces a multitude of choices across various modeling domains. For instance, in channel modeling, the choice may lie between a \textcolor{black}{high-accuracy} but computationally expensive ray-tracing model and a simpler statistical one. For user mobility, one might select from a basic random waypoint model or a sophisticated AI-based trajectory generator. Similarly, for traffic modeling, options could range from a traditional Poisson process to a complex generative model. Evaluating these components in isolation is insufficient, as their interaction and integration ultimately determine the DT's effectiveness. Task-centric DT evaluation offers a principled method for this selection. To illustrate, consider building two candidate DT environments for a resource allocation task, as shown in Fig.~\ref{fig3}. Without loss of generality, let DT~1 combine a data-driven radio map with a mathematical mobility model, while DT~2 uses a statistical channel model with a generative, NN-based user trajectory model. By sampling (state, action, reward, next-state) data from the resource allocation MDP within DT~1, DT~2, and the real network, the mismatch value against the real network for each candidate DT can be calculated \textcolor{black}{via the holistic DT evaluation framework in Section~II}. The DT with the lower mismatch value, \textcolor{black}{i.e., higher task-centric fidelity,} is provably better, as it corresponds to a tighter upper bound on the worst-case deployment loss. This guidance provides a systematic methodology for composing the most effective DT environment.

\textbf{Balancing fidelity and cost:} There is an inherent trade-off between DT's fidelity and its construction and operation cost. Modeling with higher granularity, for example, building a radio map at a 0.1-meter resolution instead of a 10-meter resolution, requires substantially more data, computation, storage, and operational resources. However, this increased investment may not translate into a meaningfully better agent policy if the targeted task is insensitive to such fine-grained details. DT evaluation via DT-BSM can be used to identify the optimal balance in this trade-off, finding the modeling granularity that minimizes cost without sacrificing training utility. To achieve this, several versions of a DT should be constructed at different fidelity levels, e.g., radio maps at 0.1-meter, 1-meter, and 10-meter resolutions. The mismatch value for each version is then calculated against the real network. As depicted in the right part of Fig.~\ref{fig3}, if decreasing the resolution from 0.1-meter to 1-meter does not cause a significant increase in the mismatch value, it indicates that the RL task's performance is robust to this level of detail. This insight allows for the deliberate selection of a model with lower granularity, reducing construction and operational costs without compromising the utility of DT and enabling a cost-effective DT tailored to a specific RL task.

\textcolor{black}{
Because our evaluation framework operates on the underlying MDP rather than specific tasks, it applies broadly to diverse 6G scenarios. In ultra-reliable low-latency communications (URLLC)~\cite{DTRL}, the reward function can be reshaped to penalize packet loss and deadline violations, enabling the task-centric fidelity metric to evaluate and select DTs that faithfully reproduce critical violation events. In AI-native RAN~\cite{AgentRAN}, DT evaluation can serve as a trustworthiness indicator for gating policy updates, ensuring that agents are trained in high-fidelity environments before physical deployment. Beyond general applicability, the framework is also compatible with existing sim-to-real techniques, which mainly focus on training robust agents to tolerate simulation errors. In contrast, the proposed evaluation proactively selects and improves the DT environment prior to policy training or deployment. Techniques like domain randomization~\cite{domainrand} or robust RL can still be applied within the selected DT to further mitigate residual domain shift, making the proposed framework a complementary extension to current sim-to-real methods.}

\section{A Real-world Case Study}
To demonstrate the practical utility of the DT evaluation framework, we provide a practical case study conducted on a real-world wireless network testbed. The experiments showcased how DT evaluation can guide the selection of high-fidelity DTs, ensuring the effectiveness of \textcolor{black}{agent} policies upon deployment.

\subsection{Experimental Setup}
\begin{figure}[!t]
\centering
\includegraphics[width=0.5\textwidth]{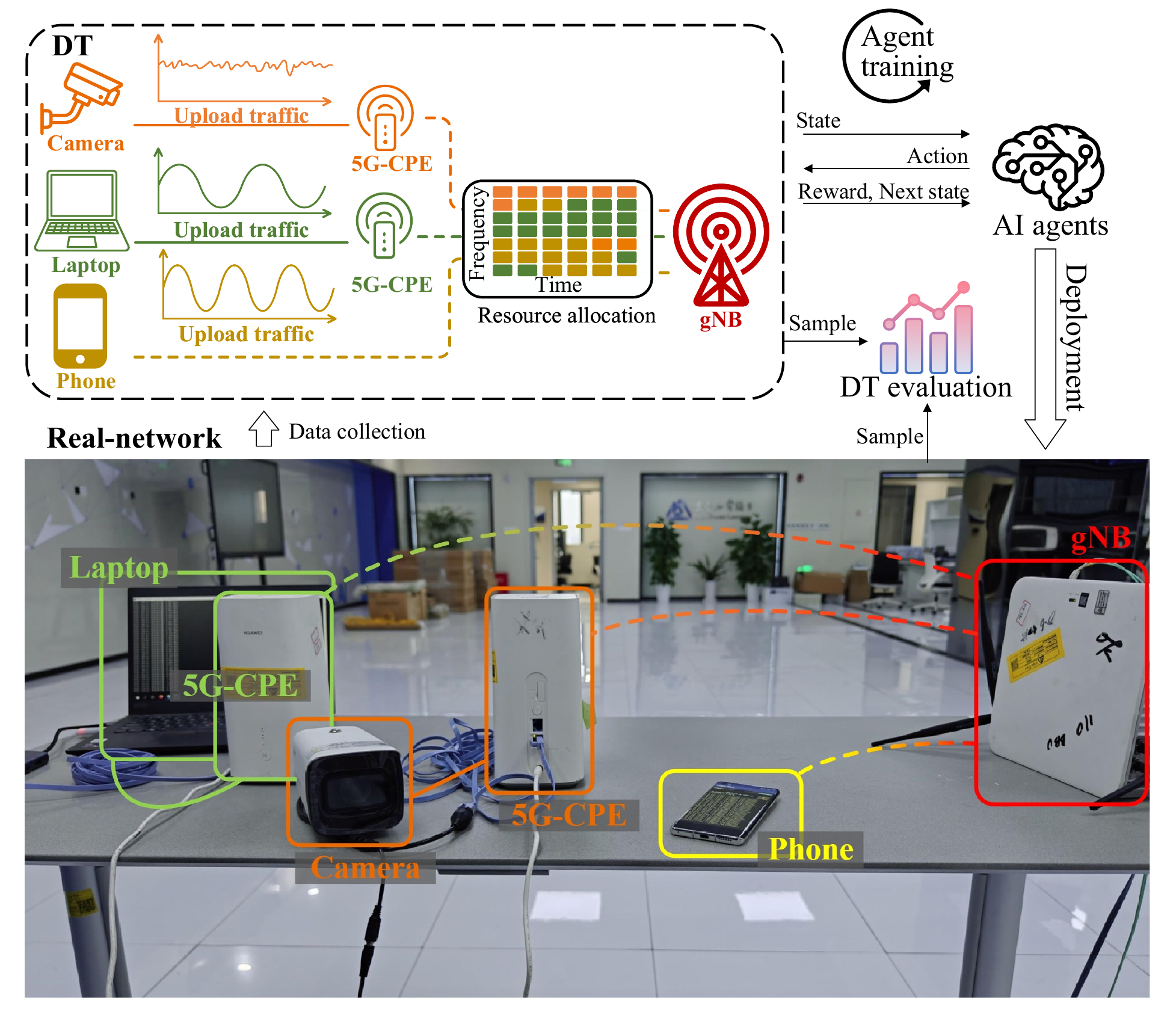}
\caption{\textcolor{black}{Case study on a real-world wireless network testbed.}}
\label{fig4}
\end{figure}

The experiment focused on a wireless resource allocation task aimed at maximizing overall user performance. The physical testbed, depicted in Fig.~\ref{fig4}, consisted of a single gNodeB (gNB) serving three distinct user equipments (UEs): a surveillance camera, a gaming laptop, and a livestreaming smartphone. The camera and laptop were connected to the network via 5G customer premise equipment (5G-CPE). This setup was designed to represent a typical environment of heterogeneous services. Each UE generated uplink traffic with unique characteristics.

\textcolor{black}{The RL agent dynamically allocates resource blocks among three UEs at each time step to maximize the weighted sum throughput. The weights are tuned to ensure all three heterogeneous services contribute comparably to the reward, thereby creating a smooth objective function to clearly demonstrate how digital twin fidelity impacts deployment performance.}
 The agent's state space includes information such as current allocated resources and throughput history, while its action space consists of possible resource block allocation schemes.

\subsection{DT Construction and Evaluation}
To create a realistic scenario for the DT evaluation, we first collected 36 hours of operational data from the live testbed, capturing a wide range of network conditions, traffic patterns, and channel dynamics. Using this dataset, we constructed a pool of 100 candidate DTs with different model settings, such as the width and depth of NN architectures and the number of training episodes, thereby yielding a collection of DTs with varying fidelity.

For each of the 100 candidate DTs, \textcolor{black}{we calculated its DT-BSM mismatch value to indicate task-centric fidelity, following the pipeline in Fig.~\ref{fig2}.} \textcolor{black}{For comparison, we also scored each DT under a model-level accuracy metric by treating the DT purely as a one-step predictor of UE traffic and computing its prediction loss against the real testbed data.} Subsequently, agents were thoroughly trained within these DTs until convergence. Each of these 100 resulting \textcolor{black}{agents} was then deployed and evaluated in the wireless network testbed to measure its final performance.

\begin{figure}[t]
\centering
\subfloat[DT evaluation result versus deployment performance]{\includegraphics[width=0.5\textwidth]{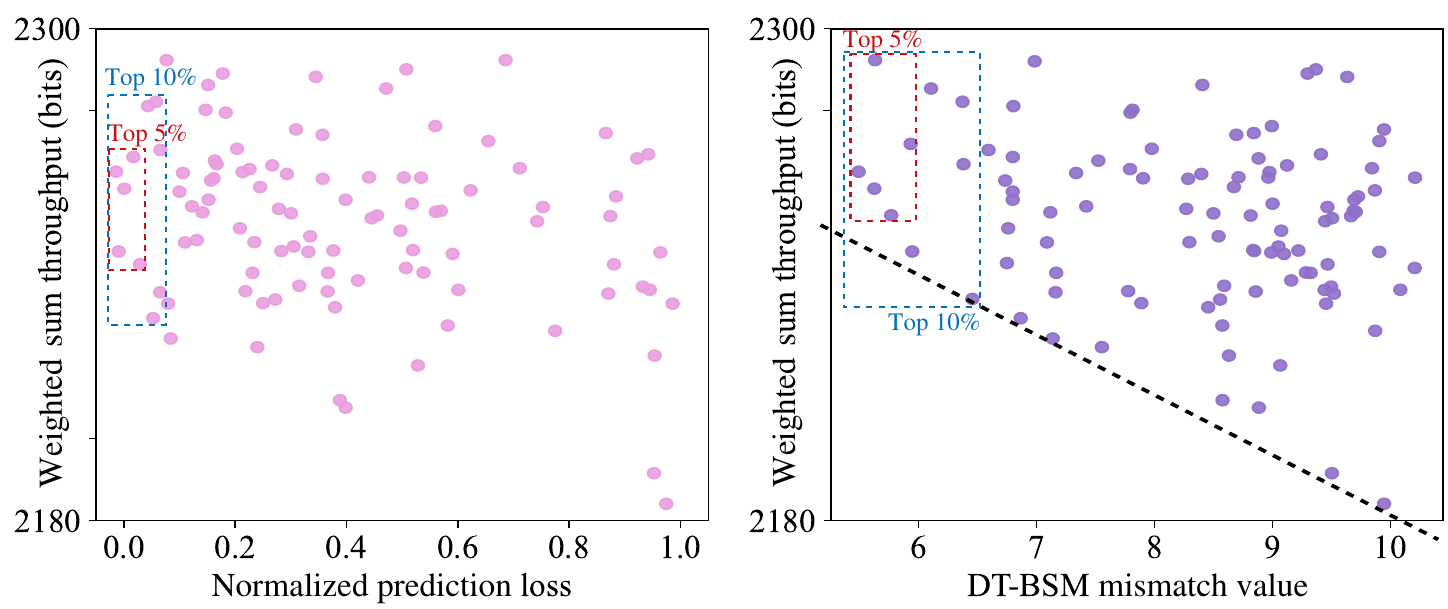}\label{Test1}}

\subfloat[Pre-filtering results]{\includegraphics[width=0.5\textwidth]{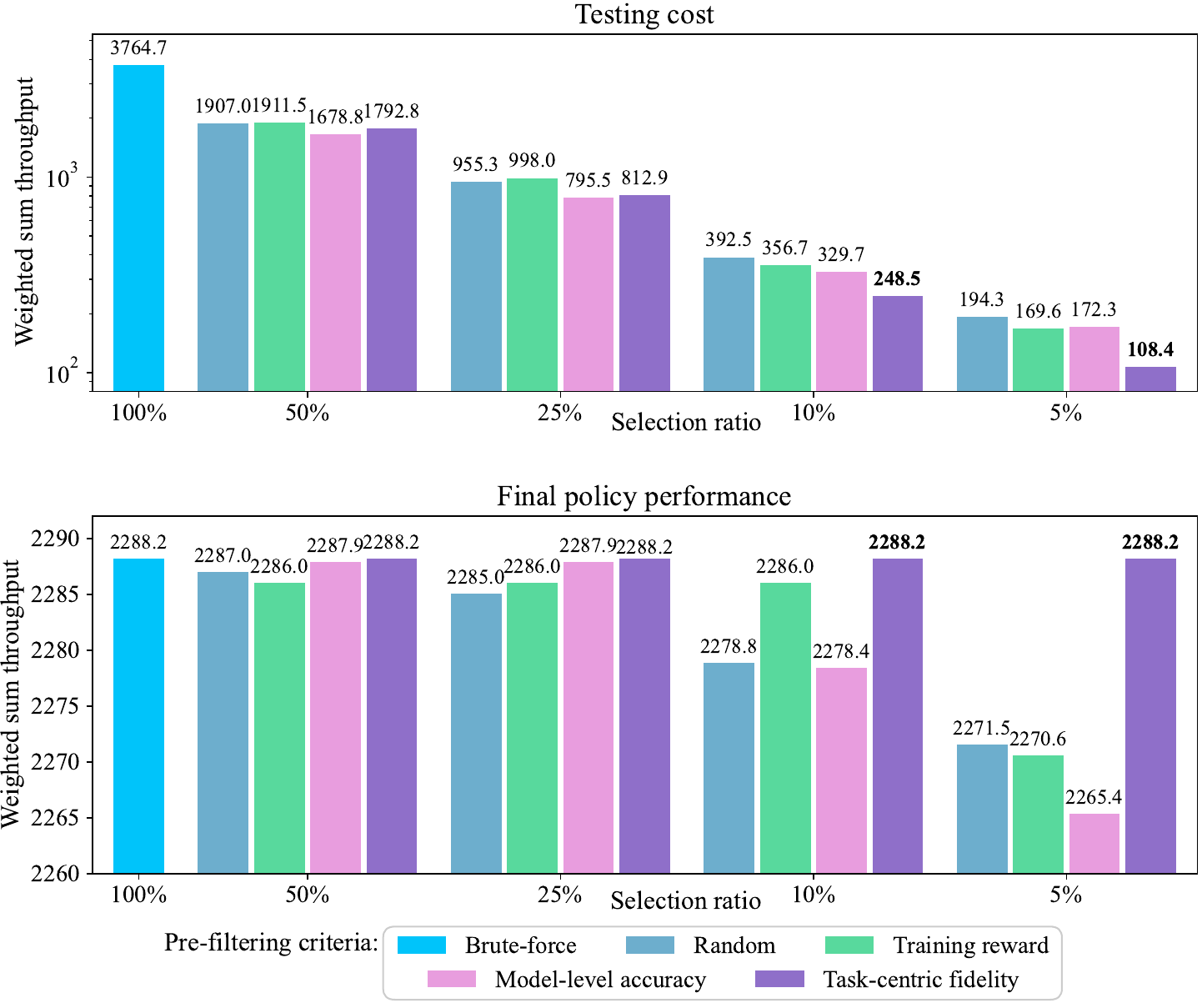}\label{Test2}}
\caption{Experimental results in a real-world case study.}
\label{Test}
\end{figure}

Fig.~\ref{Test1} empirically validates our theoretical performance bound by plotting each DT \textcolor{black}{evaluation} value against its trained policy's deployment performance. The results explicitly show a linear correlation where a lower DT-BSM consistently yields better worst-case performance\textcolor{black}{, effectively precluding training failures. This trend exactly matches the one-sided nature of the underlying bound, which controls the worst-case deployment loss for high-fidelity DTs but does not constrain what a low-fidelity DT may still produce by coincidence.} This confirms that the evaluated mismatch value reliably reflects the DT's ability to produce a robust and high-performing \textcolor{black}{agent} policy, providing a crucial performance guarantee for DT-driven network optimization. \textcolor{black}{In contrast, the normalized prediction loss exhibits a markedly weaker correlation than DT-BSM, where several DTs with low prediction loss still yield poorly performing policies, indicating that model-level accuracy has inferior discrimination ability on a DT's final deployment performance.}

\subsection{Pre-filtering}
Beyond performance guarantee, a significant practical impact of the DT evaluation lies in its ability to streamline the DT-driven RL workflow. Without a reliable evaluation method, it is necessary to train agents in all the 100 candidate DTs and then test each policy via real-world deployment to identify the best one. This exhaustive search, denoted as the \textit{brute-force} method, incurs substantial policy trial costs, encompassing both the computational resources for 100 full training processes and the testing cost of 100 separate deployments. 

The holistic DT evaluation offered an efficient alternative: pre-filtering. By calculating the mismatch value for each of the 100 candidates, we identified and selected a small subset of the most promising DTs with high \textcolor{black}{\textit{task-centric fidelity} (e.g., top 5\% as shown in Fig.~\ref{Test1})} for subsequent RL training and deployment tests. To validate its superiority, we also \textcolor{black}{introduce another three pre-filtering baselines, which rely on \textit{random} selection, the agent's \textit{training-reward} within the DT environment, and the \textit{model-level accuracy}, respectively.}

The comparison results were compelling. As depicted in Fig.~\ref{Test2}, the \textcolor{black}{task-centric fidelity}-based approach, using only the top 5\% of DTs, produced a final policy whose performance was comparable to the optimal policy found through the exhaustive brute-force method. This results in a notable reduction in resource expenditure: training costs were reduced by 95\%, and testing costs were cut by over 97\%. \textcolor{black}{Here, the testing cost is defined as the performance loss induced by using tested policies rather than the optimal policy in deployment.} In contrast, the baseline methods proved to be far less effective. While they also reduced training cost, they failed to consistently exclude the inferior DTs, leading to higher testing costs. More critically, they risked filtering out the best-performing DTs, resulting in a suboptimal final policy, particularly when the selection ratio was small.

\section{Conclusion and Future Directions}
In this article, we have provided \textcolor{black}{a holistic agent-perspective evaluation framework for wireless network DTs}. It transforms the DT into a reliable and theoretically grounded tool for \textcolor{black}{AI agent}-based wireless network optimization by \textcolor{black}{quantifying the task-centric fidelity between DT and the real world}. \textcolor{black}{We conclude by highlighting potential future directions toward more trustworthy DT-driven AI agents in 6G wireless networks.}

\subsection{Evaluation-in-the-loop Synchronization}
Current DTs are often synchronized with the real world on a fixed schedule, which can be inefficient if the system state is stable, or out-of-date if rapid changes occur. The evaluated mismatch value could serve as a dynamic trigger for DT synchronization, where the DT would be updated from the real world only when the mismatch exceeds a predetermined threshold. 
This ensures that synchronization occurs precisely when the DT begins to diverge from reality, guaranteeing both efficiency and fidelity for learning. \textcolor{black}{Meanwhile, this mechanism can be used to check whether a previous evaluation has become outdated by comparing current real-world samples against the data used in the last evaluation, which effectively addresses the non-stationary nature of wireless networks and keeps the DT trustworthy over time.}

\subsection{DT Transferability Prediction}
The cost of developing a high-fidelity DT for a complex network is substantial. A key question is whether a DT built for one environment can be effectively used in another, \textcolor{black}{especially in heterogeneous scenarios such as space-air-ground integrated networks (SAGIN), where building a high-fidelity DT from scratch for each scenario is prohibitive}. By utilizing \textcolor{black}{task-centric fidelity} to measure the similarity between two different settings, we can predict whether a DT built for one wireless network could be effectively transferred to another \textcolor{black}{while remaining reliable for agent training}, saving redevelopment costs and enabling the creation of foundational DT models that can be adapted across similar wireless network environments.

\subsection{Mismatch-aware Training}
A common challenge in DT-driven RL is that AI agents may learn to overfit to inaccuracies in the simulation. Incorporating mismatch value into the RL training enables agents that are not only optimal within the given DT but are also robust to the inherent simulation-to-reality gap\textcolor{black}{, which is particularly critical for URLLC, where overfitting to simulation can translate into unacceptable latency violations after deployment}.

\vfill

\end{document}